\documentclass[prl,aps,twocolumn,groupedaddress,showpacs,floatfix]{revtex4-1}
\pdfoutput=1
\usepackage{amsmath,amssymb}
\usepackage{graphicx}
\usepackage{bm}
\usepackage[mathscr]{eucal}
\usepackage{bbm}
\usepackage[normalem]{ulem}

\newcommand{\bs}{\boldsymbol}

\newcommand{\half}{\frac{1}{2}}
\newcommand{\td}{\text{d}}
\newcommand{\ti}{\text{i}}

\newcommand{\tx}{\tilde{x}}
\newcommand{\ty}{\tilde{y}}

\newcommand{\tkx}{\widetilde{k_x}}
\newcommand{\tqx}{\widetilde{q_x}}
\newcommand{\tqy}{\widetilde{q_y}}

\newcommand{\tX}{\tilde{Y}}

\newcommand{\tw}{\tilde{w}}

\newcommand{\cdg}{c^{\dagger}}

\newcommand{\vq}{\vec{q}}
\newcommand{\vcr}{\vec{r}}

\begin{document}
\title{Spontaneous breakdown of topological protection in two dimensions}
\author{Jianhui Wang$^1$, Yigal Meir$^{1,2}$, and Yuval Gefen$^3$}
\affiliation{$^1$Department of Physics, Ben-Gurion University of the Negev, Beer Sheva 84105, Israel}
\affiliation{$^2$Ilse Katz Institute for Nanoscale Science and Technology, Ben-Gurion University of the Negev, Beer Sheva 84105, Israel}
\affiliation{$^3$Department of Condensed Matter Physics, Weizmann Institute of Science, Rehovot 76100, Israel}
\begin{abstract}
Due to time-reversal symmetry (TRS), two dimensional topological insulators support counter-propagating helical edge modes. Here we show that, unlike the infinitely sharp edge potential utilized in traditional calculations, an experimentally more realistic smooth edge potential gives rise to edge reconstruction and, consequently, spontaneous TRS breaking. Such edge reconstruction may lead to breaking of the expected perfect conductance quantization, to a finite Hall resistance at zero magnetic field, and to a likely spin current. This calculation underpins the fragility of the topological protection in realistic systems, which is of crucial importance in proposed applications.

 \end{abstract}

\date{\today}

\pacs{73.20.At,73.21.Fg}
\maketitle

{\it Introduction.}-- The quantum spin Hall phase, a class of topological phase, was originally proposed for graphene \cite{Kane-Mele2005}, and has been subsequently understood  \cite{BHZ} to be more relevant for HgTe quantum wells, a prediction that has been later verified experimentally \cite{Konig2007}. This topological insulator (TI) phase supports helical edge modes, the only source of conduction in the system at low energies. In the quantum spin Hall phase TRS is expected to guarantee no net equilibrium charge current  at the edge. The time-reversed edge modes appear as pairs, implying zero quantum Hall conductance, but finite quantized spin-Hall conductance. Originally (the BZ model \cite{BZmodel}, describing zinc-blende materials under shear strain) this phase was treated  as a juxtaposition of two $\nu=1$ quantum Hall liquids of opposite spins with opposite directions of magnetic field. It was later realized \cite{BHZ} that such a phase may emerge in materials with intrinsic spin-orbit interactions, without an external strain, captured  by the BHZ model  \cite{BHZ} . Most importantly, these two models employed sharp boundary conditions.
Here we generalize these models to the realistic case of smooth boundary conditions, and demonstrate that TRS may be spontaneously broken, leading, for example,  to spin current and  finite Hall conductance at zero magnetic field. Importantly, this undermines the topological protection against backscattering at the edge. Moreover, we predict that the two terminal conductance through a quantum point contact  will exhibit  a conductance step  at $1\times e^2/h$,  in addition to the expected plateau  at $2e^2/h$.

The physics of this phenomenon is rather straightforward. Assume that the density of electrons is determined by an external gate. Then, in order to minimize the dominant Coulomb energy, the electron density tries to mimic the positive charge distribution on the gate. If the latter falls smoothly to zero near the edge of the system, the electron density can follow suit by separation of the edge modes, each giving rise to a decrease in density. The  smoother the confining potential, i.e. the smoother the decrease in the external positive charge, the larger the separation between the edge modes. This observation is a natural generalization of the edge reconstruction scenario predicted \cite{Dempsy93,Chamon-and-Wen,Meir,Meir96} and observed \cite{Klein1995,Zhitenev1996, Venkatachalam2012,Grivnin2014} in the quantum Hall regime. While in the integer quantum Hall regime the separation is between edge modes of the same chirality, in the present case the edge modes are of opposite chiralities, which leads to spontaneous breaking of TRS. Below we demonstrate the emergence of this phenomenon based on microscopic calculations of the two models mentioned above.

{\it The BZ Model.}-- In this model \cite{BZmodel}  the strain induced spin-orbital coupling is incorporated into the continuum limit of the conduction band, leading, for appropriate strain configurations, to an effective magnetic field in either the symmetric gauge or the Landau gauge. Hence the spectrum consists simply of the familiar Landau levels, and the single-particle wavefunctions are the Landau level wavefunctions. However, in contrast to the case of an actual external magnetic field, the direction of the effective magnetic field is opposite for the two spin species. Hence, setting the chemical potential such that only the lowest Landau level is occupied results in a pair of counterpropagating edge modes, one for each spin species. Since the two edge modes are related by time reversal transformation, elastic single-particle backscattering is forbidden unless there is explicit breaking of the TRS due to, e.g., magnetic impurities. Even though other backscattering processes such as inelastic single-particle backscattering,  two-particle backscattering, and interaction induced scattering  are allowed \cite{Wu-helical-liquid,Xu-Moore,Lezmy,Vayrynen2013,Vayrynen2014}, these processes are irrelevant in the renormalization-group sense unless the density-density interaction between the two edge modes is sufficiently strong.

The above studies did  not take into account the interplay between the confining potential and electron-electron interactions. This was addressed both in the context of the integer \cite{Chlovskii92,Dempsy93,Chamon-and-Wen} and the fractional \cite{Meir,Meir96} quantum Hall regimes. For the specific case of filling factor $\nu=1$ it was found \cite{Chamon-and-Wen}, employing the Hartree-Fock approximation, that when the confining potential is smooth enough, some electrons would detach themselves from the bulk and form an additional $\nu=1$ strip several magnetic lengths away from the main bulk of electron density. In other words, the occupation number as a function of the guiding center coordinate (i.e. the center of the single particle wavefunction in the Landau gauge) goes down from 1 to 0, then again back to 1 , then goes down to 0 and remains at 0, as one moves from the interior of the electron liquid to the exterior (cf. the red or blue curve in Fig.~1(f). Accordingly, there is now an additional pair of counterpropagating edge modes on the outer side the original one. This picture has been qualitatively confirmed by   exact diagonalisation calculations \cite{Chamon-and-Wen}. Similar physics arises, for example, at the $\nu=2/3$ fractional quantum Hall regime \cite{Meir,Meir96}, where an  additional $\nu=1/3$ strip   forms outside the minimum edge structure required by the bulk topological order \cite{Wen-Review}. This latter picture was recently used by us \cite{us} to provide a unified framework for the major experimental results in the $\nu=2/3$ fractional quantum Hall regime \cite{Chang,Bid1,Bid2},  and is also supported by  other recent experimental works \cite{Paradiso,Venkatachalam2012}. Considering the similarity between quantum Hall states and the BZ model, it is natural to ask whether a related edge reconstruction takes place within the BZ picture too.
\onecolumngrid
\begin{figure*}
\begin{center}
\includegraphics[width=.78\textwidth]{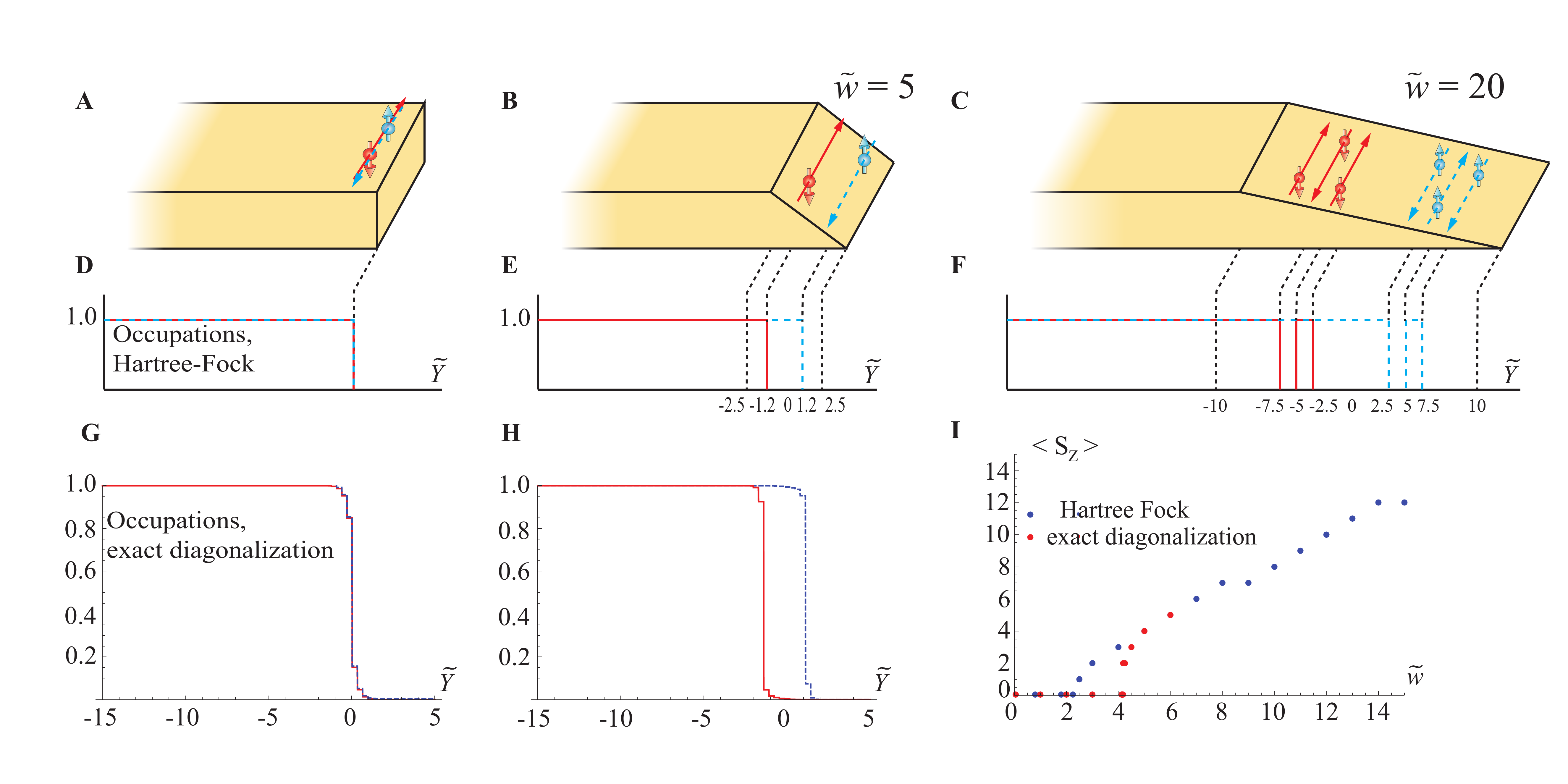}
\end{center}
\caption{\label{fig:BZ}(color online). Edge reconstruction for the BZ model. Panels (a), (b) and (c) describe the schematics of  the results for three different distributions of the confining positive charge, depicted as light orange, characterized by $\tw$, the length scale over which it decays to zero. The edge modes are marked by broken blue (spin up) and solid red (spin down) lines. Panels (d), (e), and (f) depict the occupations of the electronic states, using the Hartree-Fock approximation, demonstrating a single drop in density for a sharp edge ($\tw=0$) in (d),  spin separation for smoother edge ($\tw=5$) in (e) and Chamon-Wen-like reconstruction in (f) for even smoother edge ($\tw=20$). $\tX$ denotes the position of state,  in units of $\ell$,  the effective magnetic length, and $\tX=0$ denotes the center of the density drop. Panels (g) and (h) depict the same distributions as in (d) and (f), respectively, using exact diagonalisation. Panel (i) depicts the edge spin-magnetization as a function of the slope of the positive-charge
density, demonstrating a continuous phase transition.
}
\end{figure*}
\twocolumngrid
To check this similitude, we have studied the BZ model in the presence of a uniform positive background charge density, that decays linearly to zero at the edge of the system [Fig 1(a-c)] over a length scale $\tw$ (measured in units of the magnetic length  $\sqrt{\hbar c/eB}$, where $\hbar$ is the Planck constant, $c$ -- the velocity of light, $e$ -- the electron charge, and $B$ -- the effective magnetic field, determined in the BZ model by the applied strain). We first employed a Hartree-Fock (HF) approximation \cite{SM}, which gave rise to three different regimes. When $\tw$ is small, we find a compact spin-unpolarized ground state  [Fig.~1(d)]: as one approaches the edge, the occupation of each of the effective Landau levels drops from $1$ to $0$ at the same point (the HF analysis only allows filled or empty single-particle states). This is  the standard picture, compatible with calculations employing a sharp edge. As $\tw$ increases (smaller slope of the positive background charge), spin polarization forms [Fig.~1(e)], marking the onset of a zero temperature, continuous phase transition. This symmetry breaking is governed  by  the same physics that has been described above in the context of the quantum Hall setup - as the gradient of the positive charge background is made smaller (in absolute value), a single sharp drop in the electron density (as one moves towards the edge) would create a dipole-like charge distribution. This distribution is energetically very costly \cite{Chamon-and-Wen}. Thus, in order to minimize the electrostatic energy, a more moderate decrease of the electron density is required. Since the exchange energy is lower for compact same-spin electron density, in this regime ($\tw>\tw_{c1}$, where $\tw_{c1}\approx 2.3-2.4$) energy minimization is achieved by two consecutive steps of the electron density (one for each effective LL), at two different distances from the edge, leading to the formation of a spin polarized strip near the edge. As $\tw$ is further increased, the width of this strip and, as a consequence, the total spin polarization $S_z$, increases [Fig.~1(i)]. Once the value of  $\tw$ is sufficiently large ($\tw>\tw_{c2}$, where $\tw_{c2}\approx19-20$), the screening of the background charge by two  separate steps of the respective spin-polarized  density profiles  becomes poor, and  eventually, a Chamon-Wen type reconstruction \cite{Chamon-and-Wen} within each spin species emerges, allowing the electrons to better screen the background charge [Fig.~1(f)]. The occupation numbers and real space electronic density for each spin species are now non-monotonic, leading to the emergence of additional strips within each spin species.

To corroborate the predictions of the HF analysis, we have also performed exact diagonalisation calculations for the lower range of  values of $\tw$ \cite{SM}. For small values of $\tw$,  we have found again an  unpolarized ground-state [Fig.~1(g)]. As $\tw$ is increased, the occupancy becomes smoother, until, at larger $\tw$, we find again the emergence of spontaneous polarization [Fig.~1(h)]. The value of $S_z$ agrees with the HF prediction (for a given $\tw$), even though the critical $\tw_{c1}\approx 4.18$) is slightly larger than the HF value. To summarize, the Hartree-Fock approximation captures the essence of the problem.

\begin{figure}
\begin{center}
\includegraphics[width=.48\textwidth]{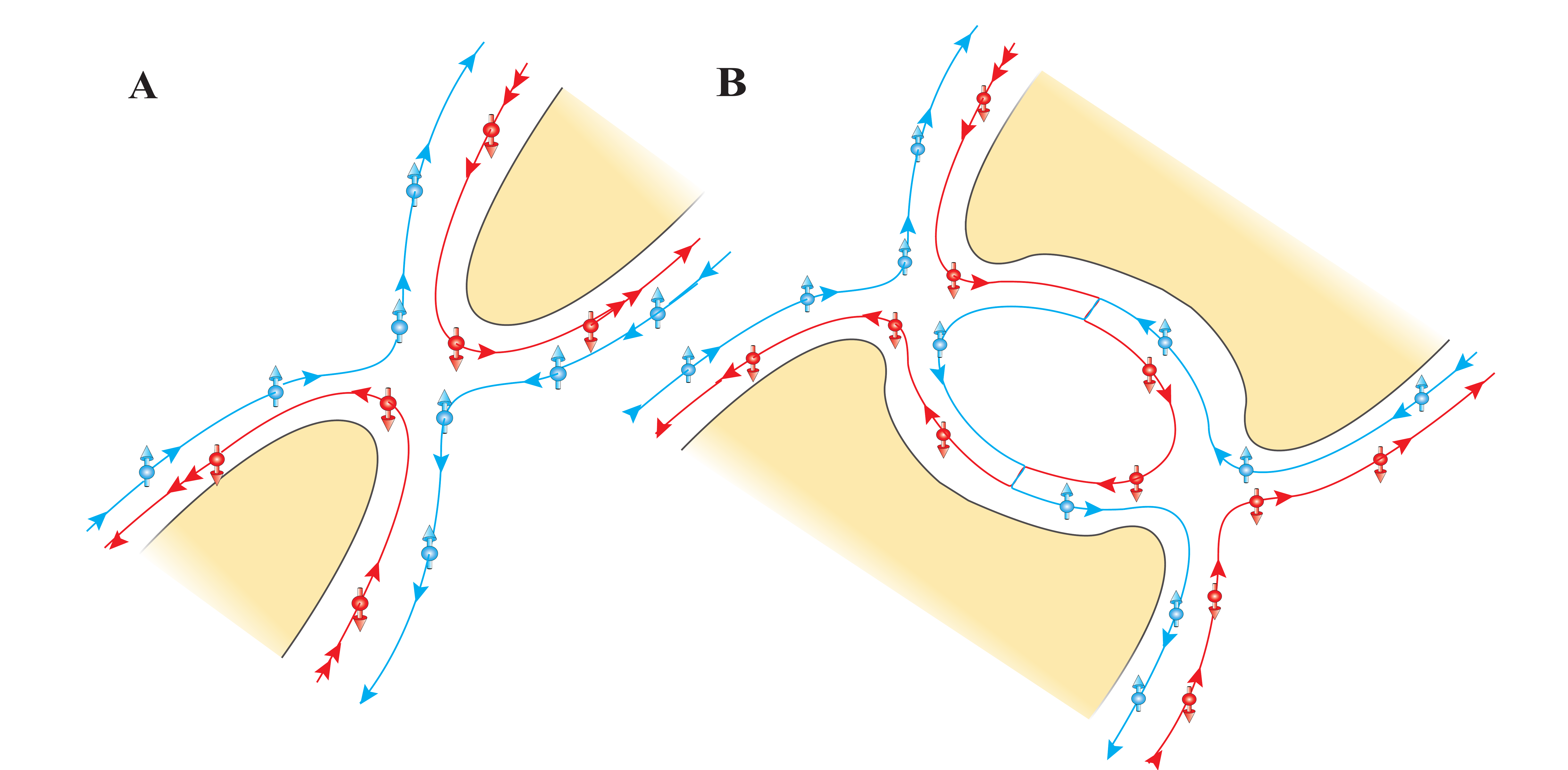}
\end{center}
\caption{\label{fig:exp}(color online). Experimental devices for detecting spontaneous spin polarization. The electrons gas is depleted in the yellow regions. Only a small part of the sample under the split gates is shown. (a) A quantum point contact may be tuned to reflect one spin channel and transmit the other, resulting in a quantum Hall effect at zero magnetic field, and spin current. (b) If the spin polarization happens to be opposite on the two sides of the two point-contact setup, a situation where each one transmit one channel, but the total transmission is zero may arise.)
}
\end{figure}

{\it Experimental Consequences.}-- The spontaneous polarization found above has important experimental implications. Since polarization means that TRS is spontaneously broken, elastic single-particle backscattering is no longer forbidden. It is well known that such a process is relevant in the renormalization group sense for repulsive density-density electron interaction and eventually leads to a localization transition when the temperature is lowered\cite{Wu-helical-liquid,Lezmy}. Owing to the spatial  separation of the two edge modes in the  spin polarized phase, the  bare backscattering amplitude should not be large. We thus expect that the violation of conductance quantization originating from  edge transport will be observed as temperature is decreased, or the sample length increases. One may thus explore the spontaneous breaking of TRS by measuring edge transport in that regime. Only at lower temperatures is localization expected to take place. This result may resolve the hitherto unexplained puzzle that transport in two-dimensional TIs appears to be ballistic as long as the samples are small\cite{Konig2007,Roth2009}, but longer samples (longer edges) exhibit lower conductance \cite{Konig2007,Knez2011,Knez2014}, providing evidence for backscattering.

An even starker demonstration of TRS breaking is provided by the setup depicted in Fig.~2(a), consisting of  a standard quantum point contact (QPC) positioned in the middle of a six-terminal device. For example, tuning the voltage on the split-gate so that the inner edge mode is completely reflected and the outer edge mode is still fully transmitted, the longitudinal and Hall resistance as functions of the split-gate voltage should have plateaus at \begin{align}
R_{xx}=&\frac{V_{23}}{I_{14}}=\frac{V_{65}}{I_{14}}=\frac{2h}{3e^2},\\
R_{xy}=&\frac{V_{62}}{I_{14}}=\frac{V_{53}}{I_{14}}=\frac{h}{3e^2},
\end{align}
respectively (for the derivation and full notation see \cite{SM}). In the general case, where the transmission probability of the outer (inner) channel is $T_O (T_I)$, $R_{xx}=(3-T_O-T_I)/(3T_I+3T_O-4T_IT_O)\ h/e^2$ and $R_{xy}=(T_O-T_I)/(3T_I+3T_O-4T_IT_O)\ h/e^2$ (the sign of $R_{xy}$ depends on the way TRS is spontaneously broken). The two-terminal conductance, given by $G_{2terminal}= (T_O + T_I)\ e^2/h$, exhibits conductance steps at $2\times e^2/h, 1\times e^2/h, 0$.  Similar steps in the conductance were observed in the quantum Hall effects\cite{Chang,Bid1}, and were regarded as evidence for edge reconstruction \cite{Beenakker,us}.  We thus predict a finite quantized  Hall resistance at zero magnetic field due to the spontaneous breaking of TRS.  Another important consequence of the breaking of TRS is the possible generation of spin-current. As the spin is a good quantum number in each of the edge modes, then reflection of one edge mode necessarily means that only one spin direction is transmitted. The transmitted edge mode can be utilized at another point in the device as a source for a specific spin. As mentioned above, in the absence of external TRS breaking, there is equal probability for the spin current to be in either direction. To tune the spin current in a specific direction,  a weak external magnetic field in the desired spin direction may be applied.

Since there is {\it a priori} full symmetry between the two edge modes, the following intriguing situation may arise when two such QPCs are put in series. Assume that the  spontaneous  symmetry breaking is different on the two sides of the sample (left and right in Fig.~2(b). Thus the two edge channels have to cross each other, on the top and bottom edges of the sample. Such a crossing defines a domain wall between a spin up and a spin down regions; it costs exchange energy, hence will  probably happen only once for each edge. Thus, if this crossing occurs somewhere between the two QPCs, the conductance through each QPC is finite ($=1\times e^2/h$), while the conductance through the two QPCs in series vanishes.

\begin{figure}
\begin{center}
\includegraphics[width=0.6\textwidth]{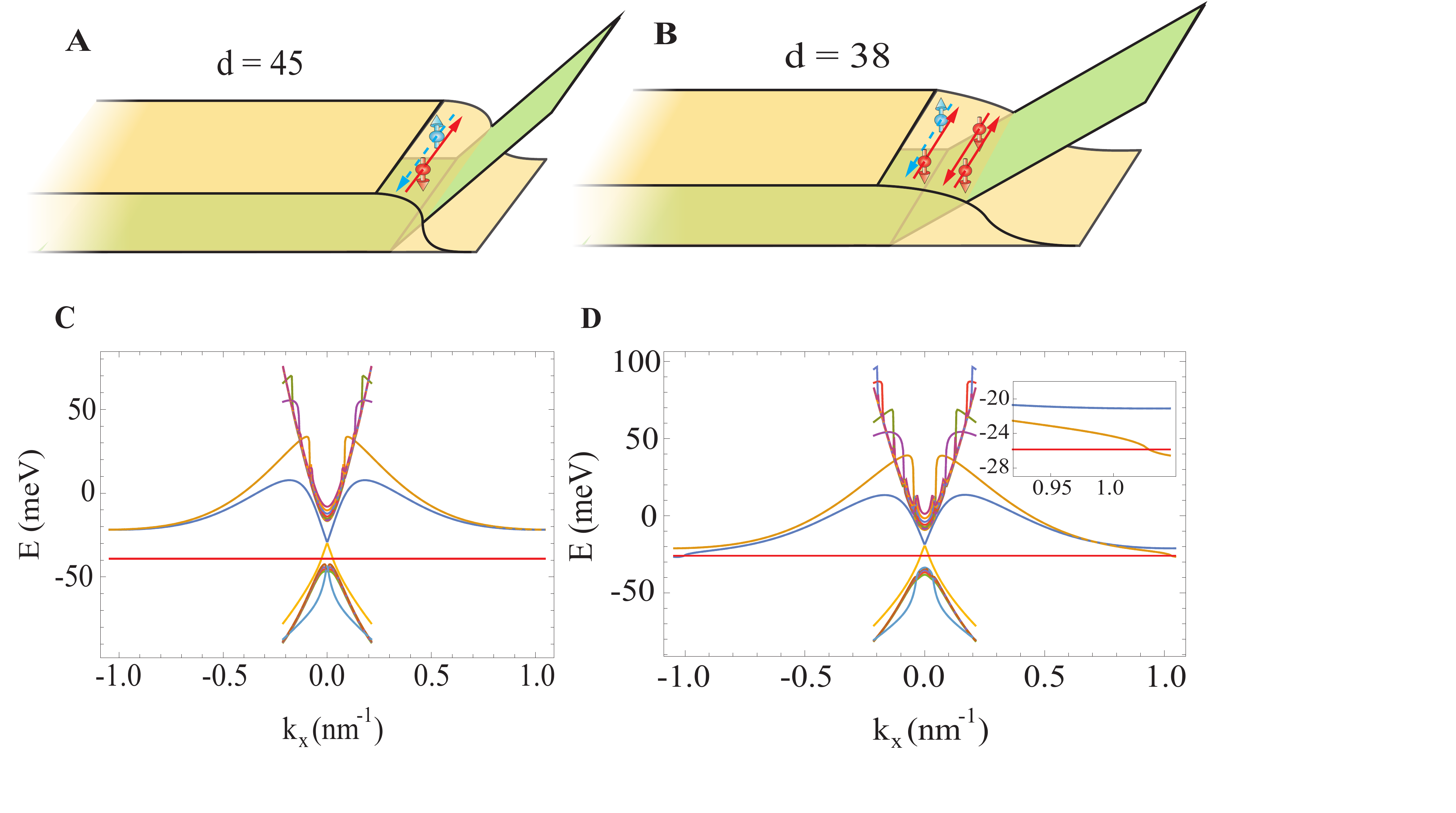}
\end{center}
\caption{\label{fig:BHZ}(color online). Edge reconstruction in the Bernevig-Huse-Zhang model. (a) and (b): Schematic presentation of the reconstruction as a function of the slope of the linear confining potential. For a sharp potential (a) each edge supports a single pair of edge modes of opposite spin and opposite direction. For a smoother edge (b) an additional pair of same spin, counter-propagating edges states emerge near the edge. Panels (c) and (d) depict the spectra for these two cases, where for the sharp potential the chemical potential intersects the spectrum at two energies, while for the smoother potential, two additional crossings give rise to two additional edge modes.
}
\end{figure}

{\it The BHZ model.}-- The BZ model has been originally developed to predict and describe topologically insulating behavior in zinc-blende semiconductors such as GaAs. Presently, however, the experimentally most studied two-dimensional TI is mercury telluride (HgTe) quantum well, which is more faithfully described by the BHZ model \cite{BHZ}. This model employs the relevant band structure that leads to band inversion \cite{Volkov1985} at specific well thickness, and nontrivial topological order.
In order to check the relevance of the physical picture described above to the BHZ model, we have performed HF calculations for this model in the presence of a linear confining potential at the edge \cite{SM}. As with the BZ model, and as found in previous studies, when the confining potential is steep (i.e. the single particle wavefunction vanishes at the edge but there is no other external potential), there is no spin polarization or edge reconstruction  [Fig.~3(a)] - there is a single pair of counter-propagating, opposite spin edge modes. Fig.~3c shows the corresponding spectrum: the chemical potential intersects the spectrum for each spin at two points, leading to two edge modes on the two edges of the system. However, as  the confining potential becomes shallower (13meV/nm for the set of parameters used in our simulations, taken from Ref.\cite{jpsj}) we observe edge reconstruction [Fig.~3(b)] -- a new pair of counter-propagating edge modes emerges near the edge. Interestingly, unlike the original counter-propagating edge modes which are of opposite spins, the reconstructed counter-propagating ones are of the same spin. The emergence of a pair of counter-propagating same-spin edge modes is very similar to the Chamon-Wen reconstruction \cite{Chamon-and-Wen}. The fact that the lowest energy state displays edge reconstruction only in one of  the spin channels is due to the gain in  exchange energy. The corresponding spectrum for this spin species [Fig.~3(d)] shows two additional crossings of the chemical potential with the spectrum. Thus, as with the BZ model, a smoother edge leads to spontaneous breaking of TRS.  This symmetry breaking facilitates  backscattering between  counter-propagating opposite-spin edge state, leading to increase in the resistance.  Moreover, the resulting difference in spin density profiles near the edge  leads to different tunneling amplitudes of the two spin directions at a quantum point contact. This yields, again in agreement with the results of the BZ model:  (i) a finite spin current  and a finite Hall conductance at zero magnetic field, as well as (ii) possible transmission blockade through two QPCs in series [Fig.~2(b)].

{\it Discussion.}-- The analysis presented here shows that a realistic smooth edge potential may lead to spontaneous breaking of TRS as well as to the appearance of additional edge modes in two-dimensional topological insulators. The broken TRS undermines topological protection and gives rise to a finite backscattering length scale,  $\ell_{BS}$. For relatively short samples, of length $L\lesssim\ell_{BS}$, the resistance will be quantized.. Introducing a QPC leads to a quantized Hall resistance at zero magnetic field (one  mode is perfectly reflected, while the other is perfectly transmitted [Fig. 2(a) and Eqs. (1) and (2)]). This is accompanied by quantized steps of the two-terminal conductance, as the QPC is pinched off. Moreover, an intricate transmission blockade through two QPCs in series may arise [Fig. 2(b)].   For longer systems $L> \ell_{BS}$ , backscattering results in deviations from conductance quantization, possibly in line with the observation of a higher resistance in HgTe            quantum wells \cite{Konig2007,Knez2011,Knez2014}. On even longer length scales,  $L>> \ell_{BS}$ , backscattering may lead to Anderson localization at the edge. These intriguing predictions, including the emergence of net spin current at zero magnetic field, are amenable to experimental test.

The applicability of the present study goes much further than TIs. It includes the quantum spin Hall effect in graphene, subject to strong magnetic field \cite{Young2014}. The fact that edge reconstruction has been predicted \cite{Meir,Meir96} and observed \cite{Venkatachalam2012,Grivnin2014} also in the fractional quantum Hall regime suggests extensions to fractional TIs \cite{Levin2009,Levin2012}. It would be intriguing to explore the additional effects due to the fractional charge of the reconstructed edge state, and whether, e.g., one can also generate, in this case, neutral edge modes in TIs \cite{KFP,Bid2}.
The present study may also put severe constraints on possible utilization of Tis, from spintronics to quantum computation \cite{Hasan-Kane-review}. As demonstrated here, smooth edge is detrimental to topological protection. For example, employing the appropriate parameters for HgTe \cite{jpsj}, we find that at the edge reconstruction transition, the bulk electron density drops to zero at the edge on a scale of 10nm. The scale implied by electrical gates is much larger, implying that we are always in the edge-reconstructed regime, unless special care is taken in defining the edge or the contacts of the device.

\begin{acknowledgments}
JW acknowledges support provided by the Kreitman Foundation. YM acknowledges ISF grant 292/15. YG acknowledges support from ISF grant 1349/14, DFG grant RO 2247/8-1, CRC 183 of the DFG,  the IMOS Israel-Russia program, and Minerva. YG acknowledges
discussions with I. Burmistrov, I. Gornyi, and
A. Mirlin.
\end{acknowledgments}
\clearpage
\setcounter{equation}{0}
\setcounter{table}{0}
\setcounter{figure}{0}

\begin{widetext}
\begin{section}{Supplemental material}

\begin{subsection}{Methods for The BZ Model}

To study edge reconstruction in the BZ model, we consider a system that is periodic in the $x$ direction and semi-infinite in the $y$ direction, such that for a sharp edge, the system extends from $y\rightarrow-\infty$ to $y=0$, and the confining potential is positioned near $y=0$. The magnetic field is applied in the z-direction and is of opposite sign for the two spin directions, $\sigma=\uparrow,\downarrow$. Following Ref.~\onlinecite{Chamon-and-Wen-S}, we model the confining potential using a positive background charge density, whose bulk value is determined by the electron bulk filling factor, and decreases linearly to zero over a width $\tw$ at the edge [Fig.~1(a-c) in the main text]. Since in the absence of interactions and external potential all the lowest LL states, characterized by their spin  $\sigma$ and by $k_x$ -- the momentum in the $x$ direction -- are degenerate, energy dispersion is only due to the many-body Hamiltonian, which has two parts,
\begin{equation}
{\cal{H}} = \sum_{\sigma k_x}\epsilon_{\sigma k_x}\cdg_{\sigma k_x}c_{\sigma k_x}
+  \half\sum_{\sigma \sigma' \{k_{x_i}\}}V^{\sigma\sigma'}_{k_{x_1}k_{x_2}k_{x_3} k_{x_4}}\cdg_{\sigma k_{x_1}}\cdg_{\sigma' k_{x_2}}c_{\sigma' k_{x_3}}c_{\sigma k_{x_4}}  .
\end{equation}
 The first term is the single-particle term due to the interaction between electrons and the positive background,  and the second term  is the two-body term due to electron-electron Coulomb interaction. $c$ and $\cdg$ are the electron annihilation and creation operators, respectively. We only consider the case where the chemical potential lies between the lowest Landau level and the first excited Landau level (hence the bulk filling factor is $\nu=\nu_{\uparrow}+\nu_{\downarrow}=1+1=2$), and ignore Landau level mixing, meaning that only the lowest Landau level is kept in the calculation;  the Landau level index is suppressed in the above expression. The matrix element of the Coulomb interaction is given by \begin{align}
V^{\sigma\sigma'}_{k_{x1}, k_{x2}, k_{x3},  k_{x4}}=&\int \td q_y\sum_{q_x}\frac{e^2}{\sqrt{q^2+\kappa^2}}\label{V-matrix-starting-point} \\
&\times \int \td \bs{r}_1 e^{-i\bs{q}\cdot \bs{r}_1}\psi^*_{\sigma k_{x1}}(\bs{r}_1)\psi_{\sigma k_{x4}}(\bs{r}_1) \int \td \bs{r}_2 e^{i\bs{q}\cdot \bs{r}_2}\psi^*_{\sigma' k_{x2}}(\bs{r}_2)\psi_{\sigma' k_{x3}}(\bs{r}_2),\notag
\end{align}
where $e$ is the electron charge, $\kappa$ is the inverse screening length, and the sum over $k_x$ is a sum over all integers $n$, such that $q_x=2\pi/L_x n$, where $L_x$ is the length of the system in the $x$ direction. $q\equiv\sqrt{q_x^2+q_y^2}$, and the single particle wavefunction is given by \begin{equation}
\psi_{\sigma k_x}(\bs{r})=\frac{1}{\sqrt{\sqrt{\pi}\ell L_x}} e^{i\tkx \tx}e^{-\half (\ty-\tX)^2}.
\end{equation}
Here  the overhead tilde indicates
dimensionless quantities defined in terms of the effective magnetic length $\ell$, e.g. $\tkx=k_x\ell$ and $\tx=x/\ell$. Note that because the two spin species experience effective magnetic fields in opposite directions, the guiding center coordinate (i.e. the center of the Gaussian part) has opposite signs (i.e. it is located on opposite edges) for a given $k_x$: $\tX\equiv -\tkx\sigma$ (here $\sigma=\pm 1$ for spin up and down), which is an important difference between the BZ model and the quantum Hall effects.

When we employ the Hartree-Fock  approximation, we assume $\langle \cdg_{\sigma k_x}c_{\sigma' k_x'}\rangle= \delta_{\sigma \sigma'}\delta_{k_x\, k_x'}n_{\sigma;k_x}$.  We do not consider HF constructions that do not have good quantum numbers specified by $H$, e.g. edge spin skyrmion. Furthermore, we assume that in the bulk $n_{\sigma \tX}=1$  (by definition, $n_{\sigma \tX}\equiv n_{\sigma;-k_x\sigma}$), and $n_{\sigma \tX}\in(0,1)$ only for $\tX$ in a finite range at the edge (we will refer to these as the active orbitals and electrons in this region as the active electrons).

The exact diagonalisation requires enumerating  all states in the relevant Hilbert space.
As this number grows exponentially as the density drops more smoothly to zero (i.e. $\tw$ increases), we were limited to $\tw\lesssim6$ in our simulations. For example, to generate Fig.~1(h) in the main text,  we used a system of 14 electrons, using 14 orbitals for each spin (so in total 28 single-particle states). The dimension of the ground state subspace is 5664.

\end{subsection}

\begin{subsection}{Method and solutions for The BHZ Model}
The BHZ calculation is done using the mapping into a fictitious tight-binding (TB) model as in  Ref.~\onlinecite{jpsj-S}. As with the BZ model, we employ periodic boundary conditions in the $x$-direction, and apply the confining potential in the $y$-direction. The many-body Hamiltonian is \begin{equation}
H=\sum_{s,\sigma}\epsilon_s \cdg_{s,\sigma}c_{s,\sigma}+\half\sum_{s_1,s_2,s_3,s_4}\sum_{\sigma,\sigma'}V^{\sigma \sigma'}_{s_1,s_2,s_3,s_4}\cdg_{s_1\sigma}\cdg_{s_2\sigma'}c_{s_3\sigma'}c_{s_4\sigma},\label{many-body-H}
\end{equation}
where $\sigma$ is the spin index, $s=(k_x,\alpha)$ is a composite index, with $k_x$ the momentum in the $x$-direction and $\alpha=1,2,\dots,2N_y$ labels the different single-particle eigenstates for that momentum and a given spin ($N_y$ is the number of rows of lattice points in the $y$ direction); $\epsilon_{s,\sigma}$ is the single-particle energy including the effects of the linear confining potential at the edge and the potential due to the ions; $\cdg$ and $c$ are creation and annihilation operators, respectively. The interaction matrix element is given by \begin{equation}
V^{\sigma \sigma'}_{s_1,s_2,s_3,s_4}=\int\frac{\td q_y}{2\pi}\frac{1}{L_x}\sum_{q_x}V_{\vq}u_{s_1,\sigma;s_4,\sigma}(\vq)u_{s_2,\sigma';s_3,\sigma'}(-\vq)\label{expression-V-matrix-element},
\end{equation}
where the sum over $q_x$ is the same as in Eq.(\ref{V-matrix-starting-point}), and \begin{equation}
u_{s_1,\sigma;s_4,\sigma}(\vq)\equiv \sum_{\vcr,\tau} e^{-\ti \vq\cdot \vcr}\psi^*_{s_1,\sigma}(\vcr,\tau) \psi_{s_4,\sigma}(\vcr,\tau),\label{mat-element-density-operator-v1}
\end{equation}
with $\vcr=(i_x \hat{x}+i_y\hat{y})a$ ($a$ is the lattice constant and $i_x=1,2,\dots,N_x$, $i_y=1,2,\dots,N_y$ label the rows in the $x$ and $y$ directions, respectively), and $\tau=\text{E1,H1}$ labels the two orbitals at each site. The single-particle eigenstate is of the form\begin{equation}
\psi_{s,\sigma}(\vcr)=\frac{1}{\sqrt{N_x}}e^{-\ti \tkx i_x}(\vec{v}_{s,\sigma})_{i_y \tau},\label{full-sp-eigenvector}
\end{equation}
where $(\vec{v}_{s,\sigma})_{(i_y \tau)}$ is the $(i_y\tau)$-th (again a composite index) component of $\vec{v}_{s,\sigma}$, the latter being a normalized eigenvector after the dependence on $i_x$ has been taken care of by the plane wave factor; and $\tkx=k_x a$ is the dimensionless wavevector. Substituting Eq.~(\ref{full-sp-eigenvector}) into Eq.~(\ref{mat-element-density-operator-v1}), performing the sum over $i_x$, we get\begin{equation}
u_{s_1,\sigma;s_4,\sigma}(\vq)=\sum_{n=-\infty}^{+\infty}\delta_{\tqx-\tkx_1+\tkx_4,2\pi n}\sum_{i_y,\tau}e^{-\ti \tqy i_y}(\vec{v}_{s_1,\sigma}^*)_{i_y \tau}(\vec{v}_{s_4,\sigma})_{i_y \tau}.
\end{equation}
Substituting this into Eq.~(\ref{expression-V-matrix-element}) and performing the $q_y$ integration, we get \begin{equation}
V^{\sigma \sigma'}_{s_1s_2s_3s_4}=\frac{e^2}{\epsilon L_x}\sum_{m=-\infty}^{+\infty}\delta_{\tkx_1+\tkx_2,\tkx_3+\tkx_4+2\pi m}\sum_{i_y,\tau}\sum_{i_y',\tau'}(\vec{v}_{s_1,\sigma}^*)_{i_y \tau}(\vec{v}_{s_4,\sigma})_{i_y \tau} [\mathbb{K}(k_{x1}-k_{x4})]_{i_y,i_y'} (\vec{v}_{s_2,\sigma}^*)_{i_y' \tau'}(\vec{v}_{s_3,\sigma})_{i_y' \tau'},\label{expression-V-matrix-element-final}
\end{equation}
where $\epsilon$ is the dielectric constant, $L_x=N_x a$ is the length of the system in the $x$ direction (the direction with the periodic boundary condition), and
\begin{equation}
[\mathbb{K}(k_{x1}-k_{x4})]_{i_y,i_y'}=\sum_{n=-\infty}^{+\infty}2K_0\left(|(\tkx_1-\tkx_4+2\pi n)(i_y-i_y')|\right)
\end{equation}
($K_0$ is the modified Bessel function of the second kind). Substituting Eq.~(\ref{expression-V-matrix-element-final}) and the usual Hartree-Fock factorization \begin{equation}
c^{\dagger}_{s_1\sigma }c^{\dagger}_{s_2\sigma'}c_{s_3\sigma'}c_{s_4\sigma}\rightarrow c^{\dagger}_{s_1\sigma}c_{s_4\sigma}\langle c^{\dagger}_{s_2\sigma'}c_{s_3\sigma'}\rangle+\langle c^{\dagger}_{s_1\sigma}c_{s_4\sigma}\rangle c^{\dagger}_{s_2\sigma'}c_{s_3\sigma'}-c^{\dagger}_{s_1\sigma}c_{s_3\sigma'}\langle c^{\dagger}_{s_2\sigma'}c_{s_4\sigma}\rangle-\langle c^{\dagger}_{s_1\sigma}c_{s_3\sigma'}\rangle c^{\dagger}_{s_2\sigma'}c_{s_4\sigma}
\end{equation}
into Eq.~(\ref{many-body-H}) and assuming $\langle c^{\dagger}_{s'\sigma'}c_{s\sigma}\rangle=\delta_{\sigma\sigma'}\delta_{ss'}\langle n_{s\sigma}\rangle$ ($n_{s\sigma}\equiv c^{\dagger}_{s\sigma}c_{s\sigma}$ is the number operator), the Hartree-Fock Hamiltonian is \begin{equation}
H_{\text{HF}}=\sum_{s,\sigma}\epsilon_{s,\sigma}\cdg_{s,\sigma}c_{s,\sigma}+\sum_{s,\sigma}n_{s,\sigma}(\sum_{s'\sigma'}\langle n_{s'\sigma'}\rangle V_{\text{H}\;ss'}^{\;\;\; \sigma\sigma'}-\sum_{s'}\langle n_{s'\sigma'}\rangle V_{\text{F}\;ss'}^{\;\;\;\sigma}),
\end{equation}
where $V_{\text{H}}$ and $V_{\text{F}}$ are the matrix elements for the direct (Hartree) and exchange interactions, respectively, \begin{align}
V_{\text{H}\;ss'}^{\;\;\; \sigma\sigma'}=&\frac{e^2}{\epsilon L_x}\sum_{i_y,\tau}\sum_{i_y',\tau'}(\vec{v}^*_{s',\sigma'})_{(i_y\tau)}(\vec{v}_{s',\sigma'})_{(i_y\tau)}[\mathbb{K}(0)]_{i_y,i_y'}(\vec{v}^*_{s,\sigma})_{(i_y'\tau')}(\vec{v}_{s,\sigma})_{(i_y'\tau')},\\
V_{\text{F}\;ss'}^{\;\;\;\sigma}=&\frac{e^2}{\epsilon L_x}\sum_{i_y,\tau}\sum_{i_y',\tau'}(\vec{v}^*_{s,\sigma})_{(i_y\tau)}(\vec{v}_{s',\sigma})_{(i_y\tau)}[\mathbb{K}(k_{x3}-k_{x4})]_{i_y,i_y'}(\vec{v}^*_{s',\sigma})_{(i_y'\tau')}(\vec{v}_{s,\sigma})_{(i_y'\tau')}.
\end{align}

The parameter values in the calculation are the same as in Ref.~\onlinecite{jpsj-S} except that $M=-21$meV so that the renormalized bulk gap is about 20meV. Also the slope of the confining potential is $11.67\text{meV}/\text{nm}$.


As mentioned in the text, the HF calculation for the BHZ model reveals that when the confining potential is shallower than 13meV/nm, edge reconstruction occurs [Fig.~3(b) and (d) in the main text], leading to an additional pair of counter propagating states of the same spin. If the the two edges of the systems are uncoupled, then we expect 4 degenerate ground states, corresponding to all combinations of the spins of the additional modes.  In our calculation, done in a finite system, we find  nearly degenerate groundstates (the difference in HF energy is $0.21$meV when the separation between the edges are about 300nm, and decreases to $0.17$meV when the edge separation is about 480nm). One solution has overall spin polarization and there are, say, three spin up edge modes and one spin down edge mode on each edge (and a degenerate state of three spin down edge modes in each edge). The second solution does not have overall spin polarization, and has, say, three spin up edge modes and one spin down edge mode on one edge, and the opposite on the other edge (with another degenerate state with opposite spins). In both cases,  on each edge there is spin polarization and a difference in the numbers of spin up and spin down edge modes, and the only difference between the solutions is the relative spin of the majority states on each edge, and thus depend on the interactions between the edges, which decays with the width the system.

\begin{figure}
\begin{center}
\includegraphics[width=\textwidth]{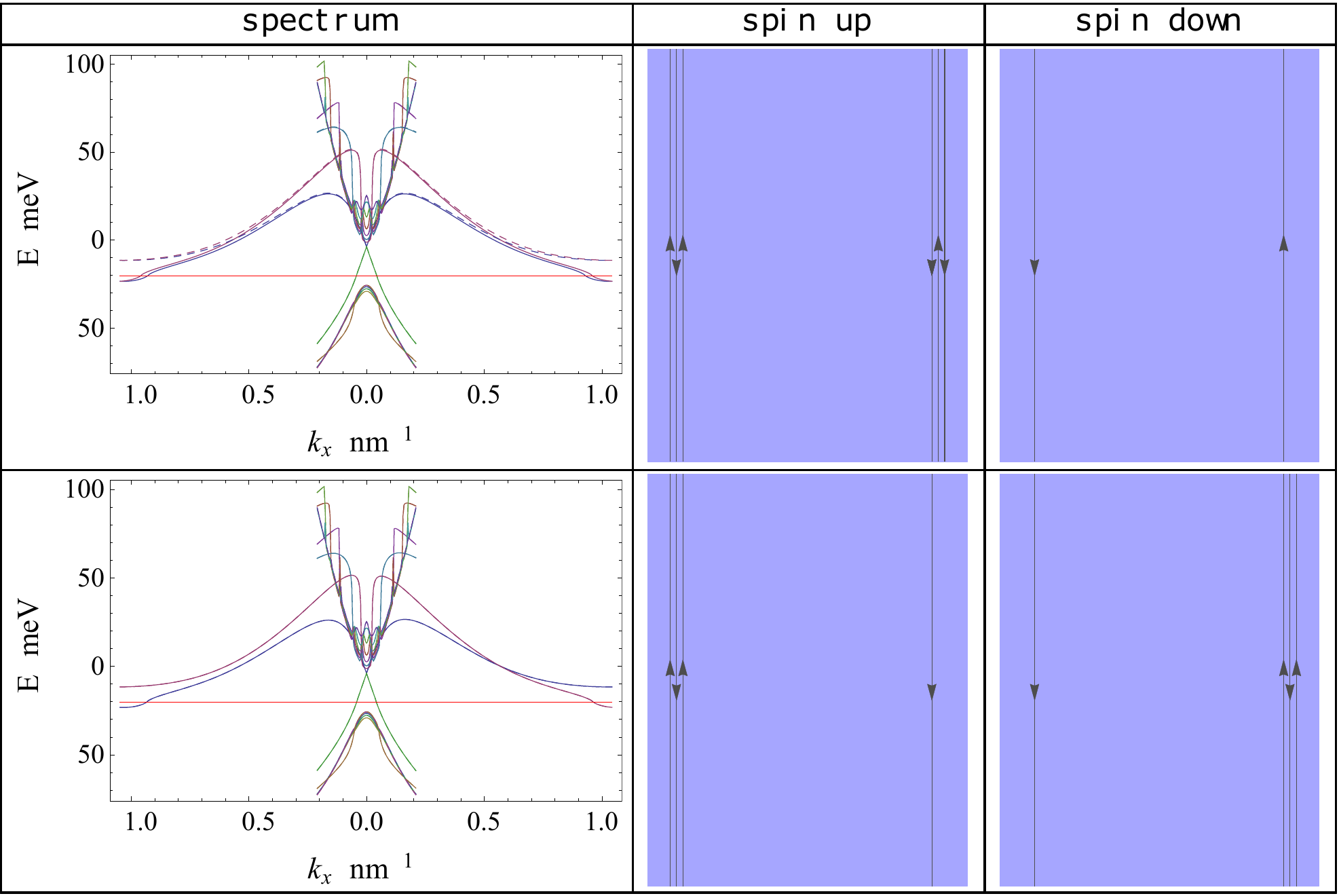}
\end{center}
\caption{\label{fig:HFforBHZ} The two solutions of the HF calculation for the BHZ model. The first row shows the result for the groundstate, the second row a near-degenerate state. The first column shows the single-particle spectrum in the Hartree-Fock approximation (only for active orbitals in the calculation). The solid lines are for spin up electrons, the dashed lines spin down electrons (in the second row the solid and dashed lines overlap). The second column shows the edge modes for the spin up electrons and the third column those for spin down electrons (as before there is the trivial degeneracy connected with flipping all the spins; and the spatial separations between the modes on a given edge is introduced for the sake of clarity). The arrows indicate the directions of the edge modes. The parameter values are those in ref.~\onlinecite{jpsj-S} except that $M=-21$meV so that the renormalized bulk gap is about 20meV. Also the slope of the confining potential is $11.67\text{meV}/\text{nm}$.
}
\end{figure}

Fig.~\ref{fig:HFforBHZ} depicts the the spectrum of the spin up and spin down electrons, and the schematics of the edge modes for the two solutions. The first row corresponds to the polarized solution, there are 3  spin up modes in each edge. Thus the chemical potential crossed the spin up spectrum (solid lines) in 6 energies, each corresponding to one edge mode. The spectrum for the spin down electrons (dashed lines) is different, and the chemical potential crossed only two energy states, similar to the non-reconstructed case [see Fig.~3(c) in the main text]. On the other hand, the second solution is unpolarized and has opposite spin configuration on the two edges. The spectrum for the two spins is now identical (second row).

Interestingly, the Hamiltonian is invariant with respect to the time reversal operator  $\hat{T}$ and reflection about a point at the center of the sample $\hat{P}$.  The ground state for the sharp edge, which supports two counter-propagating opposite spin modes on each edge, conserves  these two symmetries. Reconstruction implies broken symmetries, i.e., the symmetry classification of the Hamiltonian does not apply any more to the symmetries satisfied by the ground state. Specifically, the  $\hat{T}$ symmetry is broken. Of the possible ground states there is a class  (cf. upper row in Fig.~\ref{fig:HFforBHZ}) that conserves only $\hat{P}$ symmetry, while the second class  (lower row)   conserves only $\hat{P}\hat{T}$ symmetry. In more general terms, our analysis suggests that symmetry classification of Hamiltonians may be quite different from symmetry classification of (spontaneous symmetry broken) ground states.

\end{subsection}

\begin{subsection}{Derivation of the longitudinal and Hall resistances}
We employ a standard setup used to calculate (or measure) the longitudinal and the Hall resistances ($R_{xx}$ and $R_{xy}$, respectively), depicted in Fig.~\ref{fig:Setup}.
\begin{figure}
\begin{center}
\includegraphics[width=0.5\textwidth]{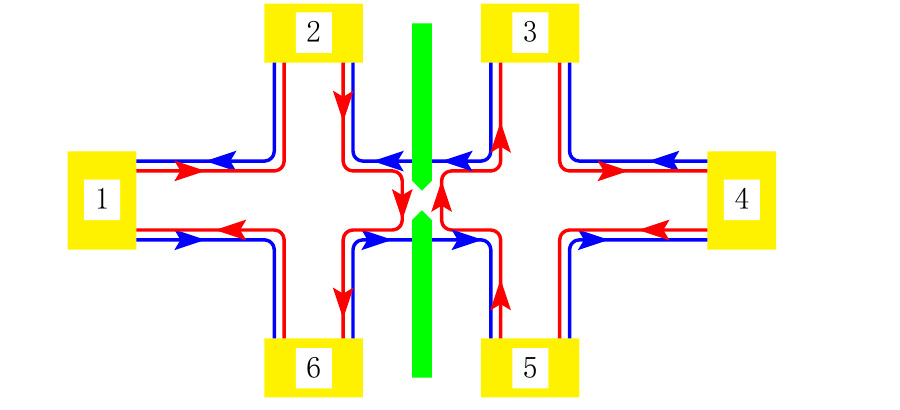}
\includegraphics[width=0.5\textwidth]{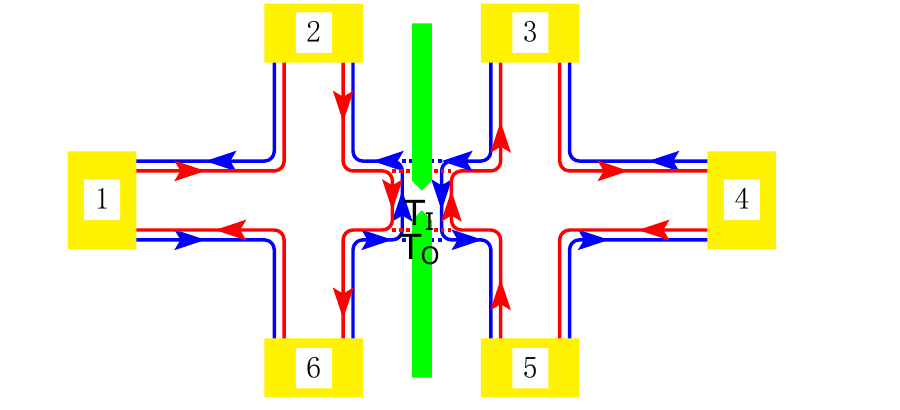}
\end{center}
\caption{\label{fig:Setup}A standard setup used to calculate the longitudinal and Hall resistances. Due to spontaneous breaking of time-reversal symmetry, we allow for different tunneling amplitudes of the two counter-propagating  spin directions. (a) Full deflection of the inner mode, full transmission of the outer mode. (b) The general case,  the inner (outer) mode has transmission amplitude $T_I (T_O)$.
}
\end{figure}
Following  B\"uttiker equations \cite{Buttiker88-S}, the resulting equations are \begin{align}
I_1=&\frac{e^2}{h}(2V_1-V_2-V_6),\\
I_2=&\frac{e^2}{h}(2V_2-V_1-V_3),\label{eq-I2}\\
I_3=&\frac{e^2}{h}(2V_3-V_5-V_4),\label{eq-I3}\\
I_4=&\frac{e^2}{h}(2V_4-V_3-V_5),\\
I_5=&\frac{e^2}{h}(2V_5-V_6-V_4),\label{eq-I5}\\
I_6=&\frac{e^2}{h}(2V_6-V_2-V_1),\label{eq-I6}
\end{align}
where $I_i$ is the current flowing out of the $i$-th terminal, $V_i$ is the voltage of the $i$-th terminal. Note that the 2nd term in the 3rd equation is $V_5$ instead of $V_2$ because of the full deflection of the inner edge mode. Similarly the 2nd term in the 6th equation is $V_2$ instead of $V_5$. Only 5 of the 6 equations are independent because the sum of the 6 equations is automatically satisfied by virtue of current conservation. With the 5 independent equations we can solve for five unknowns, e.g. $V_i-V_1$, $i=2,3,4,5,6$, in terms of the currents. Because terminals 2,3,5,6 are voltage probes, the corresponding currents are all zero, so the results will be in terms of $I_1$ ($I_4=-I_1$). Explicitly, we have (for simplicity, we set, without loss of generality, $V_1=0$)\begin{equation}
V_2=-\frac{2I_1}{3}\left(\frac{h}{e^2}\right), V_3=-\frac{4I_1}{3}\left(\frac{h}{e^2}\right), V_4=-\frac{5I_1}{3}\left(\frac{h}{e^2}\right), V_5=-I_1\left(\frac{h}{e^2}\right),V_6=-\frac{I_1}{3}\left(\frac{h}{e^2}\right).
\end{equation}
Using these, it is easy to see that \begin{equation}
R_{xy}=\frac{V_{62}}{I_1}=\frac{1}{3}\left(\frac{h}{e^2}\right)=\frac{V_{53}}{I_1},\quad R_{xx}=\frac{V_{23}}{I_1}=\frac{V_{65}}{I_1}=\frac{2}{3}\left(\frac{h}{e^2}\right),\label{RH-Rxx-plateau}
\end{equation}
where $V_{ij}\equiv V_i-V_j$. We can also consider the more general case, where the outer mode is transmitted through the QPC with probability $T_O$, and the inner mode is transmitted with probability $T_I$. Then Eqs.(\ref{eq-I2}), (\ref{eq-I3}), (\ref{eq-I5}), (\ref{eq-I6}) are modified to \begin{align}
I_2=&\frac{e^2}{h}\left[2V_2-V_1-T_OV_3-(1-T_O)V_6\right],\\
I_3=&\frac{e^2}{h}\left[2V_3-V_4-T_IV_2-(1-T_I)V_5\right],\\
I_5=&\frac{e^2}{h}\left[2V_5-V_4-T_OV_6-(1-T_O)V_3\right],\\
I_6=&\frac{e^2}{h}\left[2V_6-V_1-T_IV_5-(1-T_I)V_2\right].
\end{align}
The solution is now \begin{gather}
V_2=-\frac{-2T_O-T_I+2T_OT_I}{-3T_O-3T_I+4T_O T_I}\left(\frac{h}{e^2}\right)I_1,\: V_3=-\frac{-3-T_O+2T_OT_I}{-3T_O-3T_I+4T_OT_I}\left(\frac{h}{e^2}\right)I_1,\:\nonumber\\
V_4=-\frac{-3-2T_O-2T_I+4T_OT_I}{-3T_O-3T_I+4T_OT_I}\left(\frac{h}{e^2}\right)I_1,
V_5=-\frac{-3-T_I+2T_Ot_)}{-3T_O-3T_I+4T_OT_I}\left(\frac{h}{e^2}\right)I_1,\:\nonumber\\
V_6=\frac{T_O+2T_I-2T_OT_I}{-3T_O-3T_I+4T_OT_I}\left(\frac{h}{e^2}\right)I_1,
\end{gather}
and \begin{equation}
R_{xy}=\frac{V_{62}}{I_1}=\frac{V_{53}}{I_1}=\frac{T_O-T_I}{3T_O+3T_I-4T_OT_I}\left(\frac{h}{e^2}\right),\quad R_{xx}=\frac{V_{23}}{I_1}=\frac{V_{65}}{I_1}=\frac{-3+T_O+T_I}{-3T_I+T_O(-3+4T_I)}\left(\frac{h}{e^2}\right).
\end{equation}
It is easy to verify that these reduce to Eq.~(\ref{RH-Rxx-plateau}) for $T_O=1$, $T_I=0$.
\end{subsection}
\end{section}

\end{widetext}
\end{document}